\documentclass[12pt]{iopart}
\pdfoutput=1 
 
\usepackage{iopams}

\usepackage{cite}
\usepackage{subcaption}
\usepackage{bm}

\usepackage[pdftex]{graphicx}
\graphicspath{ {images/} }

\begin{document}

\title{Automated quantification of one-dimensional nanostructure alignment on surfaces}

\author{Jianjin Dong and Irene A. Goldthorpe}
\address{Department of Electrical and Computer Engineering, University of Waterloo, Waterloo, Ontario, Canada}
\address{Waterloo Institute for Nanotechnology, University of Waterloo, Waterloo, Ontario, Canada}

\author{Nasser Mohieddin Abukhdeir}
\address{Department of Chemical Engineering, University of Waterloo, Waterloo, Ontario, Canada}
\address{Department of Physics, University of Waterloo, Waterloo, Ontario, Canada}
\address{Waterloo Institute for Nanotechnology, University of Waterloo, Waterloo, Ontario, Canada}
\ead{nmabukhdeir@uwaterloo.ca}

\begin{abstract}
    A method for automated quantification of the alignment of one-dimensional nanostructures from microscopy imaging is presented.
    Nanostructure alignment metrics are formulated and shown to able to rigorously quantify the orientational order of nanostructures within a two-dimensional domain (surface).
    A complementary image processing method is also presented which enables robust processing of microscopy images where overlapping nanostructures might be present.
    Scanning electron microscopy (SEM) images of nanowire-covered surfaces are analyzed using the presented methods and it is shown that past single parameter alignment metrics are insufficient for highly aligned domains.
    Through the use of multiple parameter alignment metrics, automated quantitative analysis of SEM images is shown to be possible and the alignment characteristics of different samples are able to be rigorously compared using a similarity metric.
    The results of this work provide researchers in nanoscience and nanotechnology with a rigorous method for the determination of structure/property relationships where alignment of one-dimensional nanostructures is significant.
\end{abstract}

\vspace{2pc}

\submitto{Nanotechnology}

\section{Introduction}

The application of image processing to materials and nanotechnology research has the potential to enable significant advancements, both fundamental and technological.
A key activity in these research areas is the determination of \textit{quantitative} relationships between material structure and properties, which can be enabled or augmented through the use of image processing methods.

Focusing on films and surfaces, imaging techniques have become increasingly more accurate and accessible to researchers, but suitable image processing methods have not advanced at the same pace.
When successfully applied, image processing methods have resulted in an increased understanding of experimental observations in areas including nanoscale self-assembly \cite{Harrison2000, Harrison2004, Abukhdeir2008a,Abukhdeir2011a}, nanoparticle clustering \cite{Murthy2015}, molecular topography \cite{Chen2013}, and nanorod/nanowire-coated films \cite{Hu2006,Jeon2013,Mohammadimasoudi2013,Dong2015}.
These examples also demonstrate that traditional image processing techniques alone are not sufficient and must be augmented through the identification of theoretically consistent metrics for quantification of material structure.

Films and surfaces composed of one-dimensional (1D) nanostructures -- nanowires (NWs), nanorods (NRs), and nanotubes (NTs) -- are the focus of a large sub-set of materials and nanotechnology research where image processing is gaining traction \cite{Hu2006,Jeon2013,Mohammadimasoudi2013,Dong2015}.
This area is significant in that 1D nanostructures are easily transferred onto arbitrary substrates \cite{Xia2003,Vigderman2012,Park2013,Long2012,Liu2013} and there exist a broad range of applications of these materials in electronic, optical, sensing and energy devices \cite{Xia2003,Vigderman2012,Lieber2007,Zhang2013}.
Alignment of 1D nanostructures on substrates has been demonstrated \cite{Liu2012,Lau2013} and the quality of alignment has been shown to qualitatively affect many useful material properties such as the ability to polarize light and increasing surface-enhanced Raman scattering (SERS) \cite{Feng2011,Sun2010,Park2014,Qi2012}.

\begin{figure*}[h]
    \centering
    \begin{subfigure}[b]{0.3\linewidth}
        \centering
        \includegraphics[width=\linewidth]{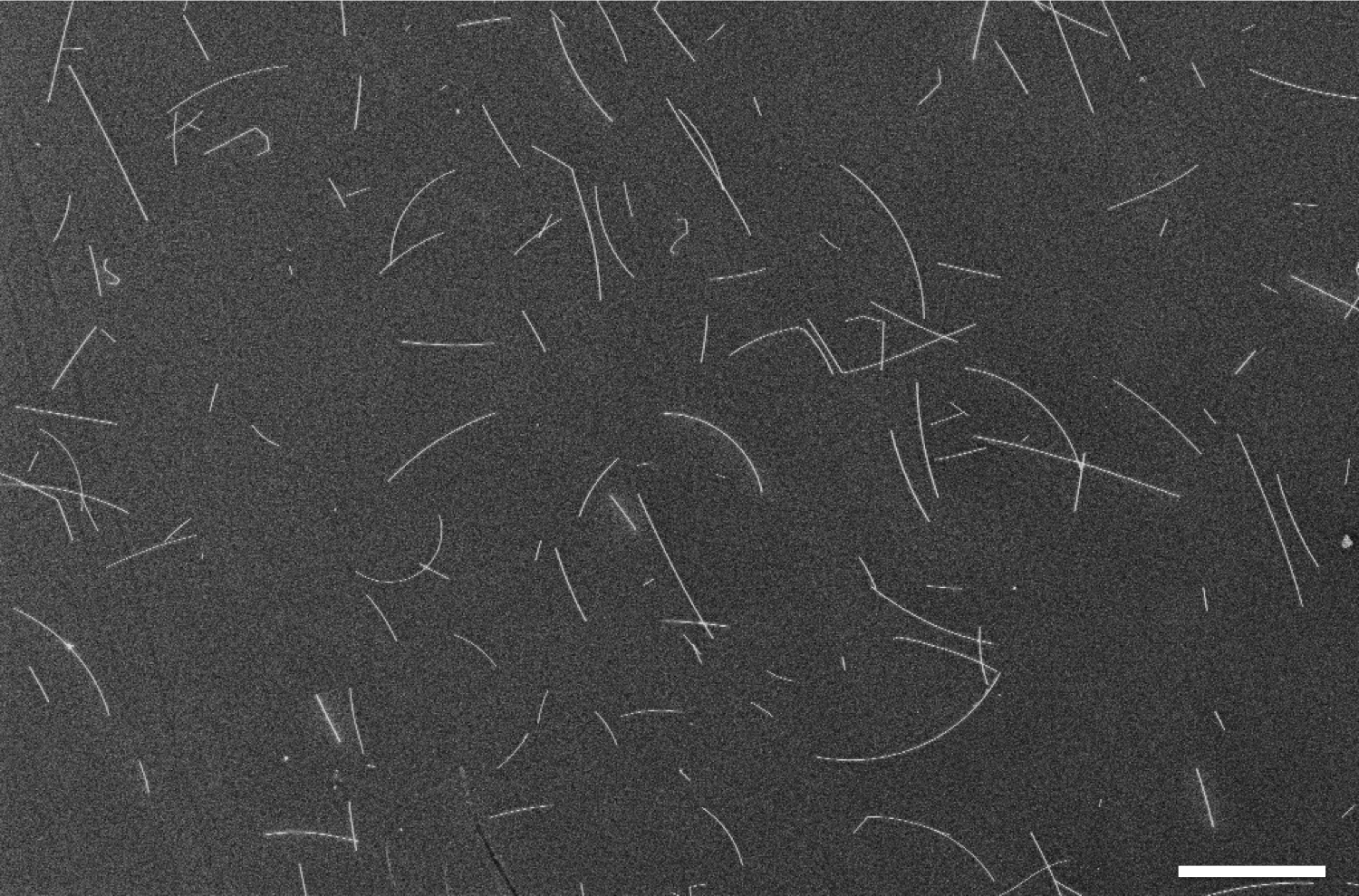}
        \caption{}\label{fig:1a}
    \end{subfigure}
    \begin{subfigure}[b]{0.3\linewidth}
        \centering
        \includegraphics[width=\linewidth]{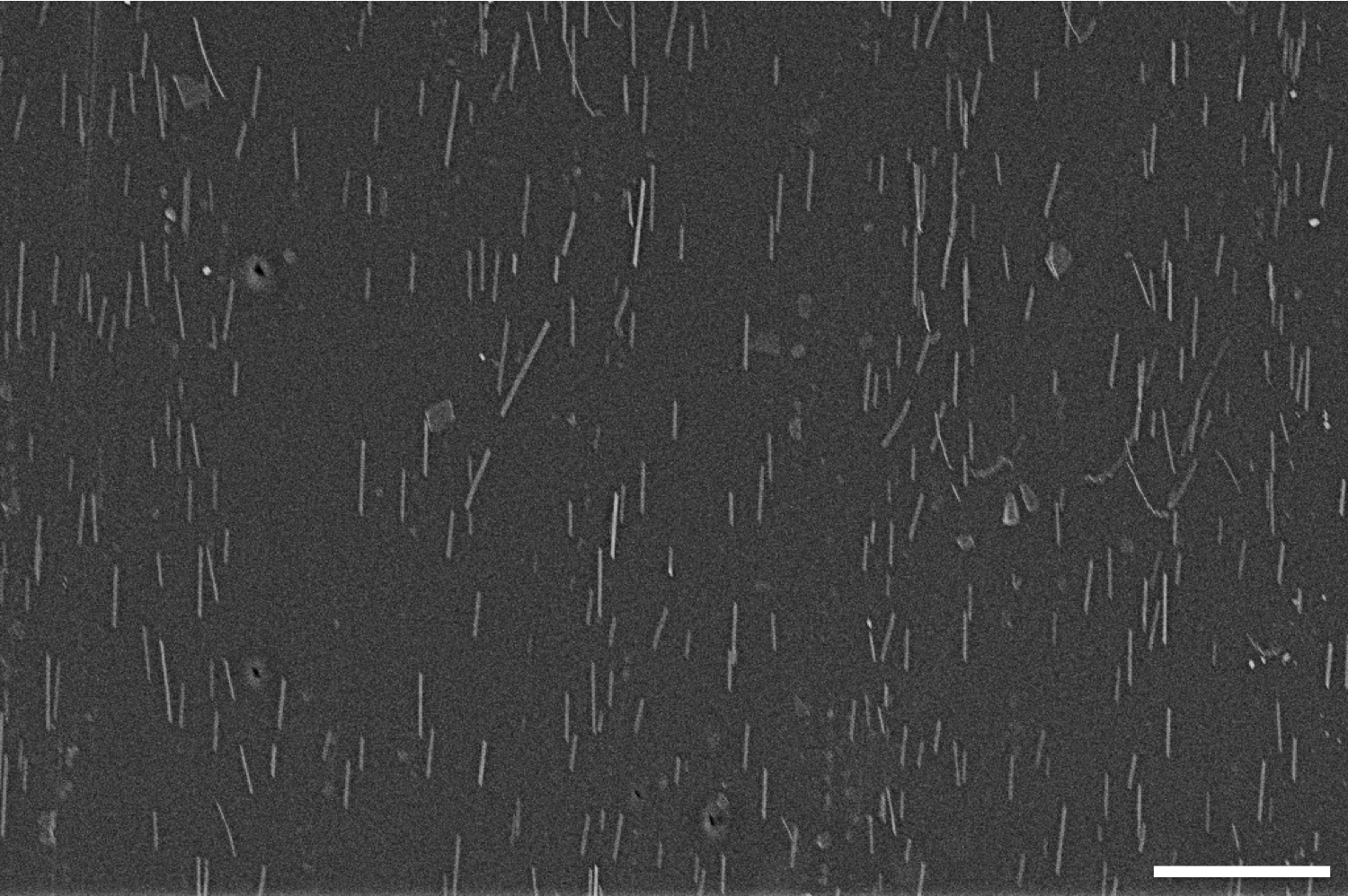}
        \caption{}\label{fig:1b}
    \end{subfigure}
    \begin{subfigure}[b]{0.3\linewidth}
        \centering
        \includegraphics[width=\linewidth]{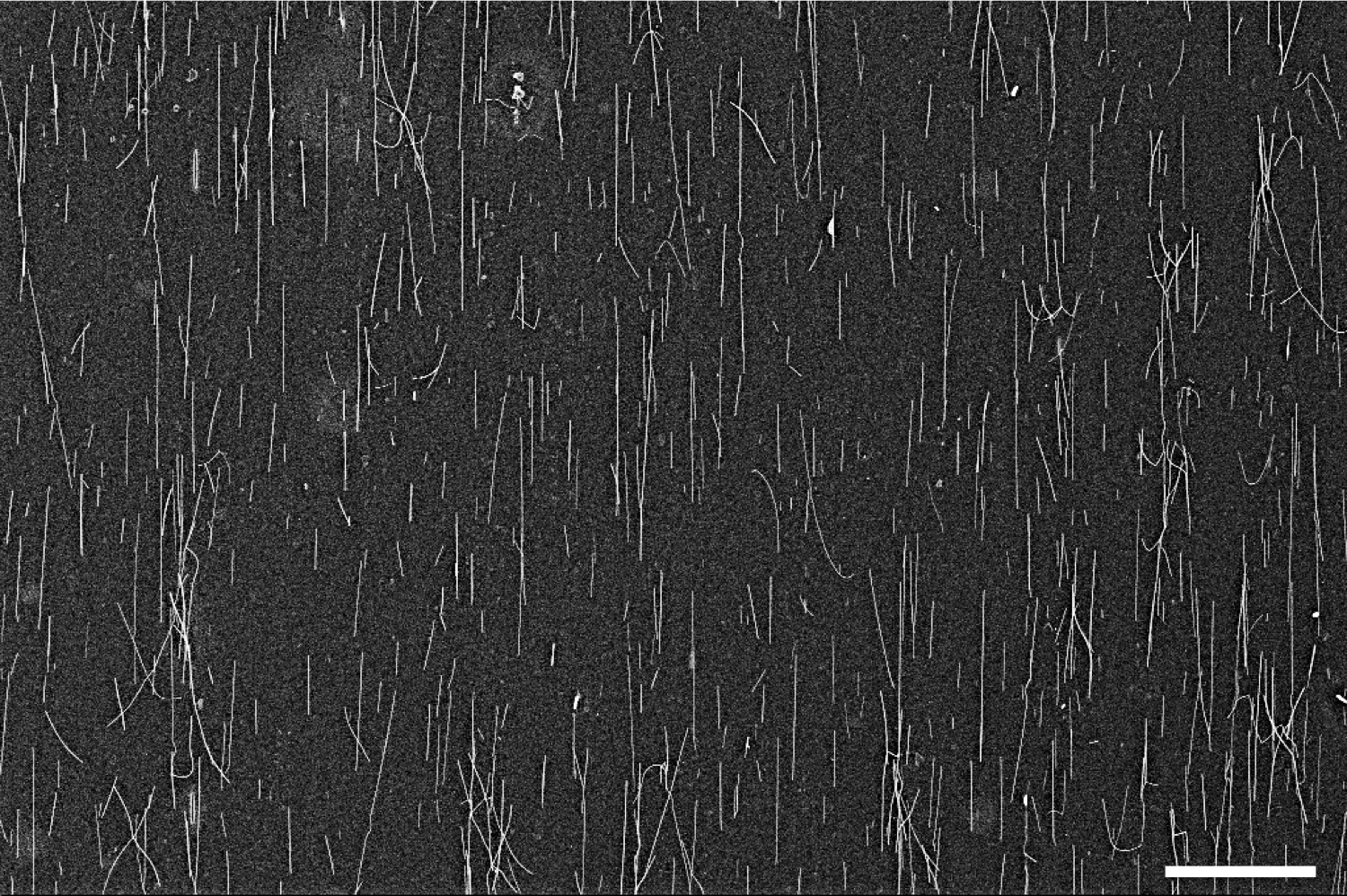}
        \caption{}\label{fig:1c}
    \end{subfigure}
    \caption{Scanning electron microscopy (SEM) images of 1D nanostructure-coated substrates with (a) sparse (non-overlapping) coverage and low alignment (scale is $50~\mu m$), (b) sparse coverage and high alignment (scale is $10~\mu m$), and (c) dense (overlapping) coverage and high alignment (scale is $40~\mu m$).} \label{fig:1}
\end{figure*}

Frequently, measurement of alignment of individual nanostructures has been done by hand through measuring the angle of alignment $\theta_{i}$ of individual 1D nanostructures \cite{Yu2007,Yao2013,Gao2015} using scanning electron microscopy (SEM) images such as in Figure \ref{fig:1}.
Not only is this approach tedious and time consuming, but it is also inaccurate and nearly impossible to execute without bias.
Automating the process through the use of image processing is clearly desirable, yet few studies have used such techniques due to the difficulty in developing suitable image processing code \cite{Hu2006,Jeon2013,Mohammadimasoudi2013,Dong2015} and the absence of suitable alignment metrics.
Recent work \cite{Hu2006} has made progress towards both quantitative and automated measurement of alignment through the use of image processing methods and the introduction of an appropriate orientational order parameter $S$:
\begin{equation}\label{eqn:order_parameter}
    S = <2 cos^{2}\theta_{i} - 1> = \frac{1}{N}\sum_{i=1}^{N} \left(2 cos^{2}\theta_{i} - 1 \right)
\end{equation}
where $\theta_{i}$ is the angle between the average alignment vector $\mathbf{n}$ and the $i^{th}$ nanostructure alignment vector $\mathbf{m}_{i}$.
$S$ values range between $0$ and $1$, with values closer to $1$ meaning the nanostructures are more aligned.
To date there has been no rigorous basis developed for alignment quantification as is present in other fields such as orientational order quantification in liquid crystalline phases \cite{Zannoni1988,deGennes1995}.
Additionally, progress has been made in imaging processing of dispersed (non-overlapping) 1D nanostructures \cite{Mohammadimasoudi2013,Dong2015}, shown in Figures \ref{fig:1a}-\ref{fig:1b}, but these approaches fail for dense 1D nanostructure coverages where overlapping is present (see Figure \ref{fig:1c}).
Thus, significant challenges still exist for automated alignment quantification and, with increasing numbers of 1D nanostructure devices and applications being developed, advances in image processing methods are as important as ever \cite{Durham2012,Duan2015,Trotsenko2015}.

The overall objective of this work is to address two of the most significant current challenges in automated alignment quantification of one-dimensional nanostructures on surfaces: (i) alignment metrics formulation and (ii) image processing of dense (overlapping) nanostructure films.
An appropriate alignment metric was first introduced in ref. \cite{Hu2006}, shown in eqn. \ref{eqn:order_parameter}.
This orientational order parameter will be shown to be a coarse approximation for the \emph{orientational distribution function} (ODF) of the nanostructures, especially for high-alignment cases.
Thus, the first objective of this work is the derivation of a complete set of order parameters which enable reconstruction of the ODF of the 1D nanostructures and, thus, rigorously quantifies alignment.
Additionally, past approaches to image processing of nanostructured films were limited to disperse non-overlapping films \cite{Hu2006,Jeon2013,Mohammadimasoudi2013,Dong2015}.
Thus, the second objective of this work is to develop an enhanced image processing method which is able to robustly and seamlessly handle both disperse (non-overlapping) and dense (overlapping) nanostructured films.

\section{Theory}\label{sec:theory}

Quantification of orientational order of materials through the introduction of appropriate orientational order parameters has been rigorously addressed in the area of liquid crystal (LC) physics \cite{deGennes1995}.
Orientational order in LC phases is traditionally quantified by a finite set of orientational order parameters \cite{Zannoni1988}:
\begin{equation}\label{eqn:3D_order_parameter}
    S_{2n} = <P_{2n} (cos\theta_{i})> 
\end{equation}
where $n$ is a positive integer, $P_{2n}$ is the Legendre polynomial of order $2n$, and $\theta_{i}$ is the angle between the average alignment vector $\mathbf{n}$ and the $i^{th}$ molecular alignment vector $\mathbf{m}_{i}$.
These order parameters were derived from a statistical mechanics representation of three-dimensional molecular alignment, the orientational distribution function (ODF) for nonpolar uniaxial molecules (in spherical coordinates):
\begin{equation}\label{eqn:odf}
    \int_{0}^{\pi} f(\theta) \sin{\theta} d\theta = 1
\end{equation}
where the ODF can be shown to have the form \cite{Zannoni1988}:
\begin{equation}\label{eqn:odf_expansion}
    f(\theta) = \frac{1}{2} + \sum_{n=1}^{\infty} \frac{4n+1}{2} S_{2n} P_{2n}(\cos{\theta})
\end{equation}
and thus a finite set of  scalar order parameters defined by eqn. \ref{eqn:3D_order_parameter} can be interpreted as a reduced-basis approximation of the exact ODF. 

While LC phases are composed of molecules whose orientation is inherently three-dimensional, nanostructures deposited on surfaces have orientation that is essentially two-dimensional.
Thus eqns. \ref{eqn:3D_order_parameter}-\ref{eqn:odf_expansion} are not applicable for the two-dimensional case.
Past research has been performed on LC phases constrained to two-dimensions in which a two-dimensional orientational order parameter $S = <\cos{2 \theta_{i}}>$ was first introduced by Straley in ref. \cite{Straley1971}.
This can be shown to be equivalent to eqn. \ref{eqn:order_parameter} using simple trigonometric identities.
For two-dimensional LC phases, the scalar order parameter was later expanded on in ref. \cite{Cuesta1990} introducing a two-dimensional alignment tensor:
\begin{equation}\label{eqn:2D_order_tensor}
    \mathbf{Q} = <2 \bm{m}_{i} \bm{m}_{i} - \bm{\delta}>
\end{equation}
which provides a simple approach to compute the average molecular alignment vector $\mathbf{n}$ through eigenvalue decomposition of $\mathbf{Q}$.

The derivation of a suitable set of two-dimensional orientational order parameters in this work closely follows that of Zannoni in ref. \cite{Zannoni1988} for the three-dimensional case.
An ODF for a set of nonpolar cylindrically symmetric objects constrained to two-dimensions must obey the following constraints:
\begin{equation}\label{eqn:2D_odf}
    f(\theta) = f(\theta + i\pi)
\end{equation}
where $i$ is an integer and the normalization condition:
\begin{equation}\label{eqn:2D_normalization}
    \int_0^{2\pi} f(\theta) d\theta = 1
\end{equation}
An appropriate orthogonal expansion for $f(\theta)$ exists in terms of a Fourier cosine series,
\begin{equation}
    f(\theta) = \frac{1}{2 \pi} + \frac{1}{\pi} \sum_{n=1}^{\infty} S_{n}\cos{n\theta}
\end{equation}
which is further constrained by eqn. \ref{eqn:2D_odf} to include terms with only even integers:
\begin{equation}\label{eqn:2D_odf}
    f(\theta) = \frac{1}{2 \pi} + \frac{1}{\pi}\sum_{n=1}^{\infty} S_{2 n}\cos{2 n\theta} 
\end{equation}
As with the three-dimensional orientational order case, this expansion defines a consistent set of orientational order parameters:
\begin{equation}\label{eqn:2D_order_parameter}
    S_{2n} = <\cos{2 n\theta_{i}}> = N^{-1}\sum_{i=1}^{N} \cos{2 n \theta_{i}} 
\end{equation}
the first of which $S_{2}$ is consistent with eqn. \ref{eqn:order_parameter}.
Through computation of these order parameters the ODF (eqn. \ref{eqn:2D_odf}) may be reconstructed with increasing accuracy as higher order $S_{2n}$ terms are included.

\section{Computational Methods}\label{sec:computational}

Image processing of microscopy images typically consists of four sequential tasks \cite{Dong2015}: filtering, thresholding, object detection, and shape fitting.
In this work, both filtering and thresholding methods are used which are essentially unchanged from past work \cite{Hu2006,Jeon2013,Mohammadimasoudi2013,Dong2015}, although they will be summarized here for clarity.
However, the presented image processing method differs substantially from past approaches in that mathematical morphology methods \cite{Haralick1987,Dougherty2003} are used for object detection and characterization, as opposed to computationally intensive shape fitting tasks used in past work \cite{Hu2006,Jeon2013,Mohammadimasoudi2013,Dong2015}.

\begin{figure*}[h]
    \centering
    \begin{subfigure}[b]{0.24\linewidth}
        \centering
        \includegraphics[width=\linewidth]{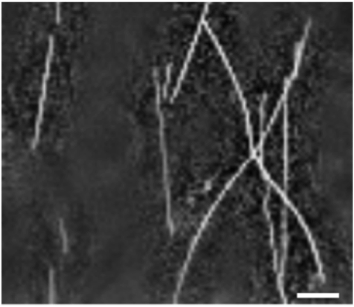}
        \caption{}\label{fig:2a}
    \end{subfigure}
    \begin{subfigure}[b]{0.24\linewidth}
        \centering
        \includegraphics[width=\linewidth]{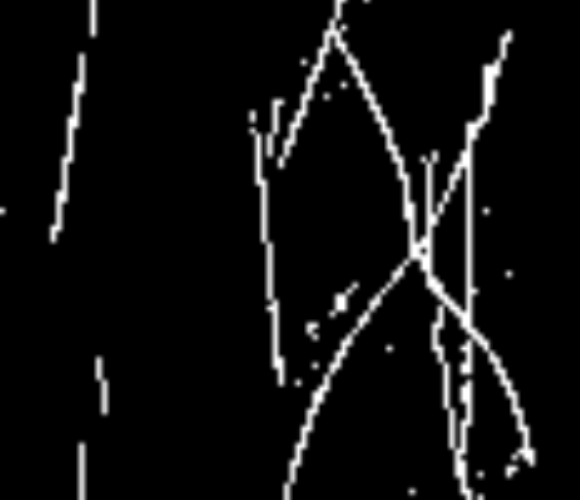}
        \caption{}\label{fig:2b}
    \end{subfigure}
    \begin{subfigure}[b]{0.24\linewidth}
        \centering
        \includegraphics[width=\linewidth]{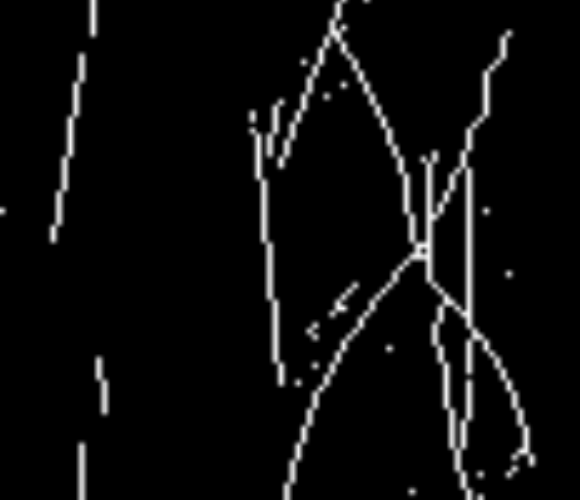}
        \caption{}\label{fig:2c}
    \end{subfigure}
    \begin{subfigure}[b]{0.24\linewidth}
        \centering
        \includegraphics[width=\linewidth]{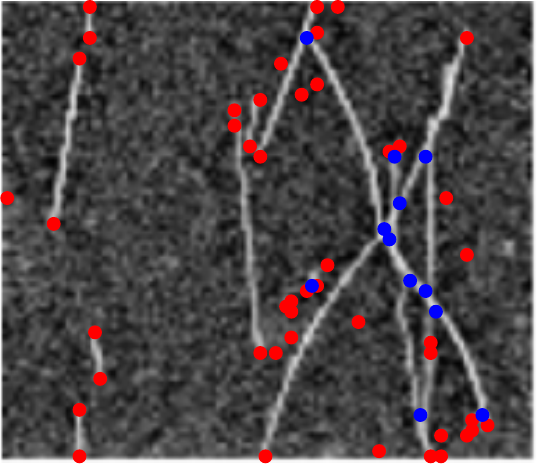}
        \caption{}\label{fig:2d}
    \end{subfigure}
    \caption{(a) A denoised sub-region of the SEM image from Figure \ref{fig:1c} (rotated, scale is $5~\mu m$), (b) the binary image resulting from thresholding of the sub-region using Otsu's Method, (c) the topological skeleton resulting from morphological analysis of the binary image, and (d) end-point and branch-point pixels resulting from further morphological analysis of the skeleton superimposed on the original image.} \label{fig:2}
\end{figure*}

The filtering and thresholding tasks are used to remove measurement noise from the raw microscopy image and segment the grayscale image into a binary image, respectively.
A non-local denoising filter \cite{Buades2011, OpenCV} is used with a length scale chosen to be smaller than the smallest characteristic nanostructure.
Depending on the variation of background intensity in the image, either Otsu's Method \cite{Otsu1979,OpenCV} or adaptive thresholding \cite{OpenCV} is used on the filtered grayscale image resulting in a binary image where each pixel is either foreground (1) or background (0).
This binary image is the starting point for automated identification of one-dimensional nanostructures.
Sample binary images resulting from filtering and thresholding of a sub-region of Figure \ref{fig:1c} are shown in Figures \ref{fig:2a} and \ref{fig:2b}, respectively.

Given a binary image, past work \cite{Hu2006,Dong2015} used least-squares fitting of ellipses to foreground objects to identify candidate one-dimensional nanostructures.
This single operation both uniquely identifies foreground objects and provides approximations of their morphological quantities such as major axis length, minor axis length, aspect ratio, eccentricity, and axes orientations.
This approach has limited applicability in that nanostructures need to be non-overlapping and thus is useful only for dispersed samples such as shown in Figures \ref{fig:1a}-\ref{fig:1b}.

In order to robustly and seamlessly process images with both non-overlapping and overlapping nanostructure films, a \emph{topological skeleton} \cite{Dougherty2003} generated from the binary image is used in the presented method.
A topological skeleton preserves foreground object shape but reduces its representation to a simple set of discretized curves which are more amenable to characterization.
There are many methods for generating topological skeletons from a binary image; in this work morphological operators are used, specifically using the \emph{hit-or-miss transform} \cite{Dougherty2003}.
This approach is chosen in that morphological operaters are also used to determine end-points and branch-points (overlapping areas) of one-dimensional nanostructures from the topological skeleton. 
Given a binary image in Figure \ref{fig:2b}, the resulting topological skeleton is shown in Figure \ref{fig:2c}.

Characterization of the one-dimensional nanostructures is facilitated by identification of end-points and branch-points in the topological skeleton.
End-points in the topological skeleton correspond to end-points of the nanostructures in the image, while branch-points correspond to overlap areas of nanostructures.
These points are efficiently identified through the use of mathematical morphology operations on the topological skeleton (Figure \ref{fig:2c}); sample output is shown in Figure \ref{fig:2d}.
For each object in the binary image, end-points and branch-points located within them may then be grouped together for further analysis (Section \ref{sec:results:image_processing}), including a second level of filtering to exclude features below a certain length threshold which results from artifacts introduced by the presence of noise in the source image.

\section{Results and Discussion}

\subsection{Dispersed (Non-overlapping) Nanowire Films}\label{sec:results:dispersed}

\begin{figure*}[h]
    \centering
    \begin{subfigure}[b]{0.3\linewidth}
        \centering
        \includegraphics[width=\linewidth]{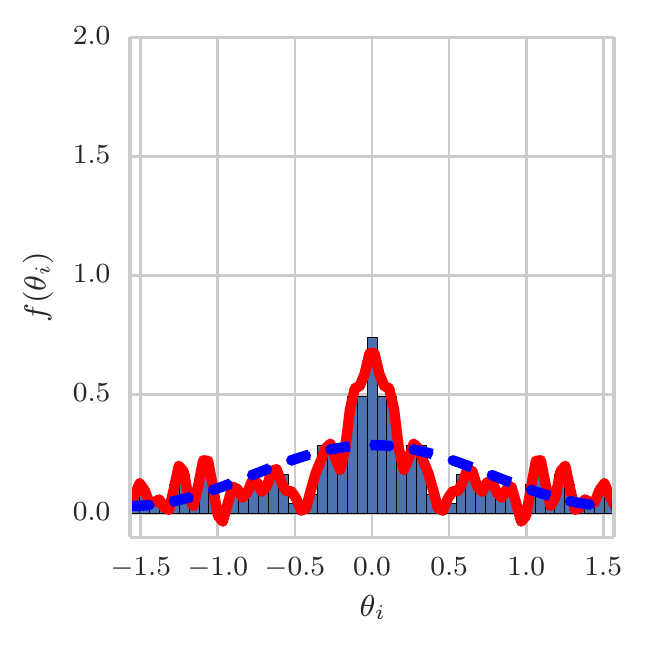}
        \caption{}\label{fig:3a}
    \end{subfigure}
    \begin{subfigure}[b]{0.3\linewidth}
        \centering
        \includegraphics[width=\linewidth]{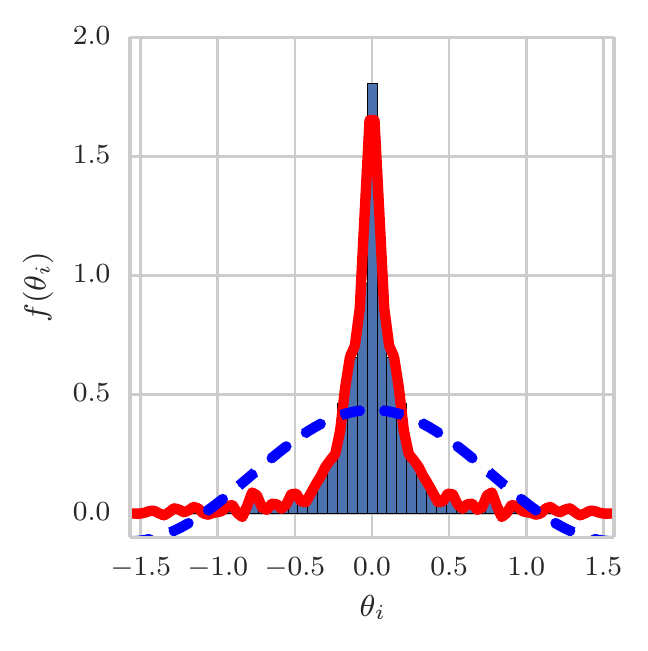}
        \caption{}\label{fig:3b}
    \end{subfigure}
    \begin{subfigure}[b]{0.3\linewidth}
        \centering
        \includegraphics[width=\linewidth]{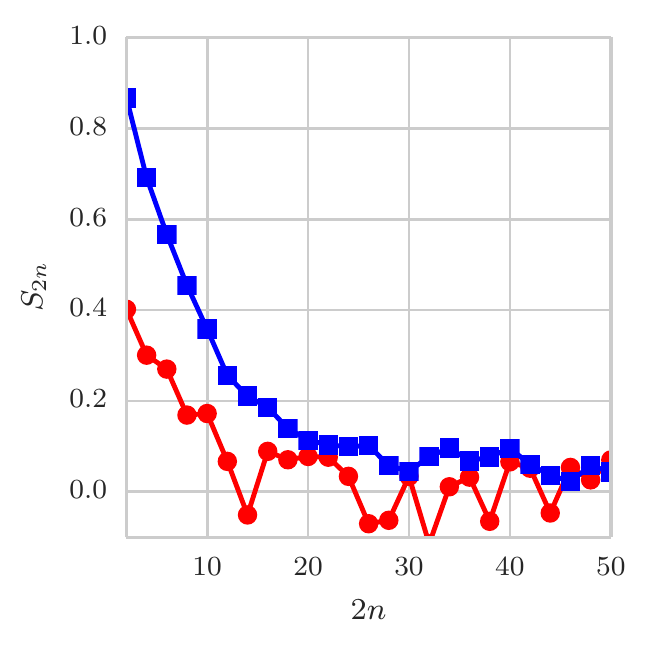}
        \caption{}\label{fig:3c}
    \end{subfigure}
    \caption{Single (dashed line) and multiple (solid line) orientational order parameter reconstructions of the ODF $f(\theta_{i})$ superimposed over histogram plots of the nanostructure orientations $\theta_{i}$ values from SEM images shown in (a) Figure \ref{fig:1a} and (b) Figure \ref{fig:1b}. (c) Plots of the magnitude of orientational order parameters $S_{2n}$ from the image in Figure \ref{fig:1a} (circles) and \ref{fig:1b} (squares).} \label{fig:3}
\end{figure*}

A distinguishing feature of 1D nanostructure-covered films is that they are able to be fabricated with an extremely high degree of alignment with $S > 0.9$ \cite{Dong2015}, as shown in Figures \ref{fig:1b}-\ref{fig:1c}.
Conversely, orientational order found in liquid crystalline materials is typically low, where $S \approx 0.3-0.6$.
In high alignment regimes, single orientational order parameter measures of alignment \cite{Hu2006} are insufficient for accurate reconstruction of the ODF, $f(\theta)$, and thus not adequate for rigorous quantification of alignment.
In order to demonstrate this, SEM images of disperse (non-overlapping) 1D nanostructure-covered films were analyzed using the image processing method presented in ref. \cite{Dong2015}.
Figure \ref{fig:1a} was used for the low alignment case and Figure \ref{fig:1b} for the high alignment case; order parameters (eqn. \ref{eqn:2D_order_parameter}) up to order $50$ were computed.
Figures \ref{fig:3a} and \ref{fig:3b} show histograms of the angle $\theta_{i}$ between the orientation of individual nanostructures and the average orientational axis.
As is seen in Figure \ref{fig:3a}, the single order parameter approximation is adequate for the low alignment case in that higher order parameters quickly approach zero, shown in Figure \ref{fig:3c}, which has oscillation of the higher order parameters resulting from the small sampling size of nanostructures in the image.
However, the single order parameter approximation fails for the high alignment case, where a relatively complex ODF is reconstructed (Figure \ref{fig:3b}) and higher order parameters are non-zero up to order $30$ (Figure \ref{fig:3c}).
Once again, oscillation in the higher order terms is observed due to the relatively small sample size.

A single orientational parameter is useful to determine if order is present and its approximate degree, but it does not reveal details of the angular distribution of the nanostructures. Two well-aligned samples with the same S could have a very different composition of nanostructure orientations, and this composition can affect the properties of the aggregate film.
An example of two images with similar $S_{2}$ values but different higher order parameters is discussed in Section \ref{sec:results:dense}.

\subsection{Optimized Image Processing Method}\label{sec:results:image_processing}

Given the combination of image segmentation and mathematical morphology image processing methods reviewed in Section \ref{sec:computational}, for a given image of 1D nanostructures, sets of end points and branch points for each contiguous feature can be computed.
For dispersed films (Figures \ref{fig:1a}-\ref{fig:1b}), 1D nanostructure orientation and length can be easily approximated from this data in that each feature should have only two end-points and no branch-points (since there are no overlapping nanostructures).
Given, for each feature, a pair of end points $\{\mathbf{r}_{1}, \mathbf{r}_{2}\}$ the nanostructure alignment vector is $\mathbf{m}_{i} = l_{i}^{-1} (\mathbf{r}_{1} - \mathbf{r}_{2})$ with the nanostructure length $l_{i} = ||\mathbf{r}_{1} - \mathbf{r}_{2} ||_{2}$.

\begin{figure*}[h]
    \centering
    \begin{subfigure}[b]{0.45\linewidth}
        \centering
        \includegraphics[width=\linewidth]{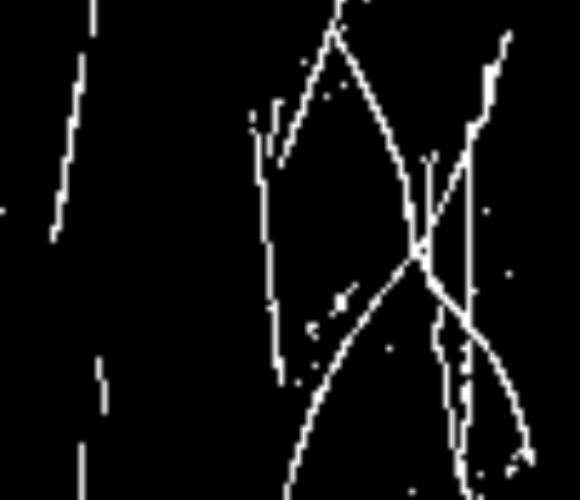}
        \caption{}\label{fig:4a}
    \end{subfigure}
    \begin{subfigure}[b]{0.45\linewidth}
        \centering
        \includegraphics[width=\linewidth]{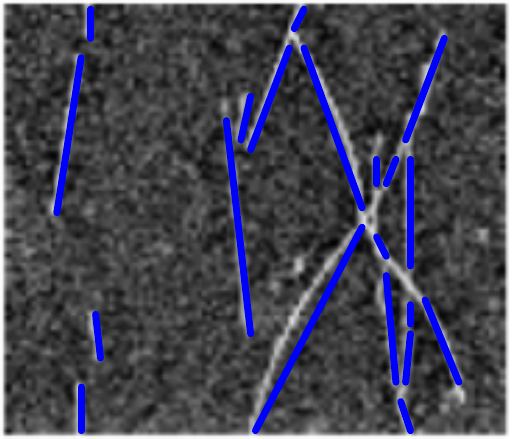}
        \caption{}\label{fig:4b}
    \end{subfigure}
    \caption{Intermediate image processing results for the SEM image shown in Figure \ref{fig:2a}: (a) the binary image resulting from applying Otsu's method and (b) the NW segments found from applying the enhanced algorithm to the topological skeleton and filtering segments below a user-specified threshold (5 pixels).} \label{fig:4}
\end{figure*}

In the frequent case of overlapping 1D nanostructures, the identification of end points for each of them becomes significantly more difficult.
Instead of developing a complex iterative method to determine end points, a more simple method is proposed.
Given the case where two or more 1D nanostructures overlap, a single feature in the binary image (Figure \ref{fig:4a}) will contain multiple 1D nanostructures and at least one branch-point will be identified within it.
The addition of an intermediate step is proposed where branch points within the topological skeleton computed from the binary image are removed and treated as if they are background pixels.
The modified topological skeleton now has no branch points and, for example, a feature which originally had two nanostructures overlapping now corresponds to four separate non-overlapping nanostructures.
The method for non-overlapping nanostructures may now be applied in that each feature in the modified topological skeleton has no branch points and only two end points.
The result of applying this image processing method to the SEM image shown in Figure \ref{fig:2a} is shown in Figure \ref{fig:4b}.

This approach does result in a loss of information in that the original topological skeleton is modified such that contiguous nanostructures are now non-contiguous.
The implication of this is that determination of original nanostructure length is not possible, but orientation is unaffected.
Within the present context of quantifying alignment of the nanostructures, this can be resolved through a reformulation of eqn. \ref{eqn:2D_order_parameter}.
Assuming that each nanostructure has an equal length $l$ and, thus, equal contribution to the orientational order of the film:
\begin{equation}
        S_{2n} =  (l N)^{-1}\sum_{i=1}^{N} l\cos{2 n \theta_{i}} = N^{-1}\sum_{i=1}^{N} \cos{2 n \theta_{i}} 
\end{equation}
Now, taking into account that each nanostructure can have different lengths $l_{i}$ a weighted average can be used:
\begin{equation}\label{eqn:length_weighted_order_parameter}
        S^{w}_{2n} =  L^{-1}\sum_{i=1}^{N} l_{i}\cos{2 n \theta_{i}}
\end{equation}
where $L=\sum_{i} l_{i}$ is the total length of nanostructures present in the image.
This approach both accounts for the differing length of nanostructures in the orientational order parameter formulation and circumvents the need for identification of a unique set of nanostructures in the image.
Any nanostructure or combination of nanostructures may be decomposed into an arbritrary set of smaller nanostructures without any affect on the order parameters computed through eqn. \ref{eqn:length_weighted_order_parameter}, unlike with eqn. \ref{eqn:2D_order_parameter}, which weights each nanostructure orientation equally.
Furthermore, even for non-overlapping samples, eqn. \ref{eqn:length_weighted_order_parameter} might be more appropriate than eqn. \ref{eqn:2D_order_parameter} if one desires the alignment of longer nanostructures to be weighted more heavily than shorter ones. 

\subsection{Dense (overlapping) Nanowire Films}\label{sec:results:dense}

\begin{figure*}[h]
    \centering
    \begin{subfigure}[b]{0.3\linewidth}
        \centering
        \includegraphics[width=\linewidth]{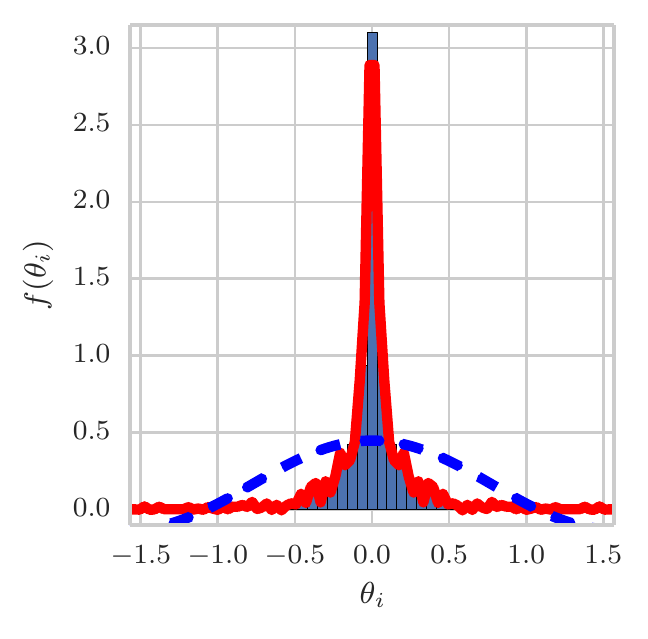}
        \caption{}\label{fig:5a}
    \end{subfigure}
    \begin{subfigure}[b]{0.3\linewidth}
        \centering
        \includegraphics[width=\linewidth]{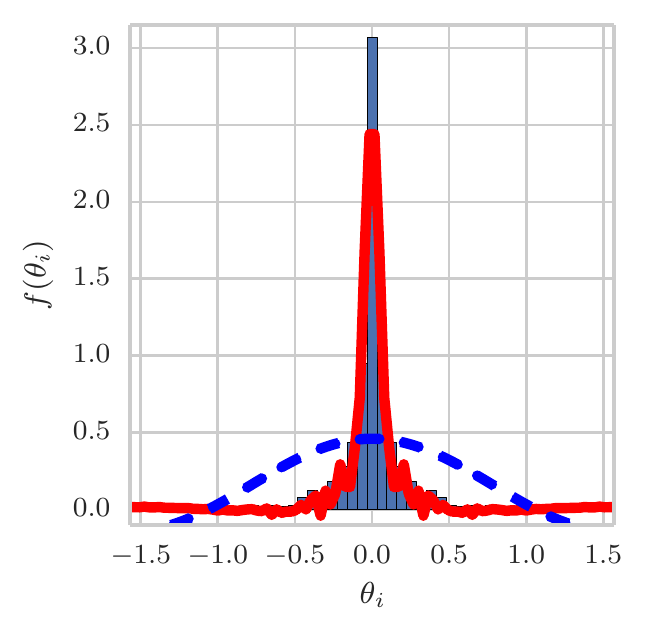}
        \caption{}\label{fig:5b}
    \end{subfigure}
    \begin{subfigure}[b]{0.3\linewidth}
        \centering
        \includegraphics[width=\linewidth]{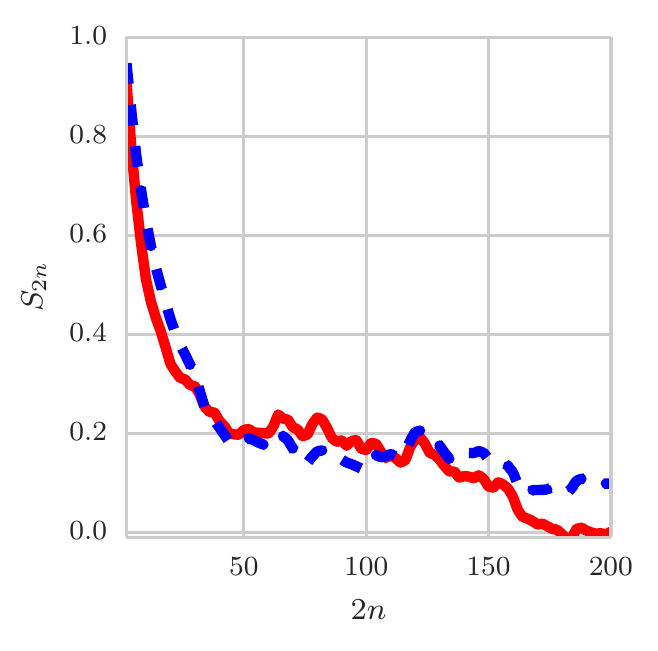}
        \caption{}\label{fig:5c}
    \end{subfigure}
    \caption{Single (dashed line) and multiple (solid line) orientational order parameter reconstructions of the ODF $f(\theta_{i})$ superimposed over histogram plots of the nanostructure orientation $\theta_{i}$ values from the SEM image shown in Figure \ref{fig:1c} using the ODF reconstruction with (a) the standard order parameters $S_{2n}$ and (b) the length-weighted order parameters $S_{2n}^{w}$. (c) Plots of the magnitude of orientational order parameters $S_{2n}$ (solid line) and $S^{w}_{2n}$ (dashed line).} \label{fig:5}
\end{figure*}

Combining the ODF and image processing methods from Sections \ref{sec:results:dispersed}-\ref{sec:results:image_processing} results in a highly robust and descriptive method for quantification of 1D nanostructure alignment.
The combined approach was applied to the dense nanostructure film shown in Figure \ref{fig:1c} in order to both demonstrate its application and compare the use of non-weighted (eqn. \ref{eqn:2D_order_parameter}) and weighted (eqn. \ref{eqn:length_weighted_order_parameter}) order parameters to reconstruct the ODF.
Figures \ref{fig:5a}-\ref{fig:5b} show reconstructions of the orientational distribution function using single and multiple order parameters for the unweighted ($S_{2n}$) and weighted ($S^{w}_{2n}$) formulations, respectively.
The nanostructures in the SEM image shown in Figure \ref{fig:1c} are highly aligned and, thus, the single order parameter approximation again is found to be insufficient in reconstructing the ODF.
Instead, orientational order parameters up to order $200$ were required to reconstruct the ODF due to the extremely high alignment, where $S_{2} \rightarrow 1$.

Comparing the multiple order parameter reconstructions, a non-negligible correction of the unweighted ODF (Figure \ref{fig:5a}) is found when compared to the more accurate weighted ODF reconstruction (Figure \ref{fig:5b}).
Figure \ref{fig:5c} shows the magnitude of orientational order parameters of increasing order, which also shows a non-negligible difference.

\begin{figure*}[h]
    \centering
    \includegraphics[width=0.45\linewidth]{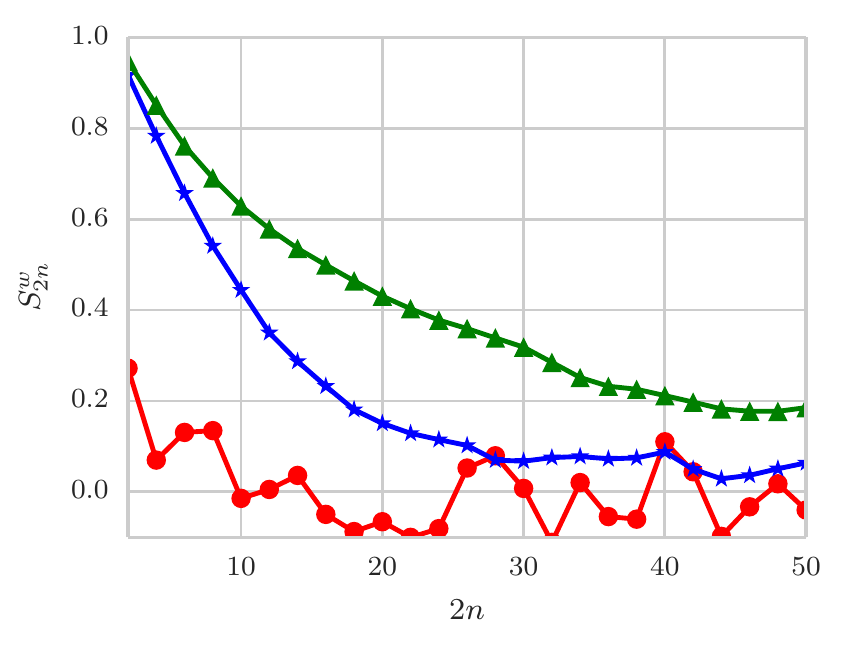}
    \caption{Plots of the magnitude of the length-weighted orientational order parameters $S^{w}_{2n}$ for the SEM images shown in Figures \ref{fig:1a} (circles), \ref{fig:1b} (stars), and \ref{fig:1c} (triangles).} \label{fig:6}
\end{figure*}

In order to further support the use of multiple orientational order parameter quantification of alignment, length weighted order parameters were also computed for the dispersed nanostructure images shown in Figures \ref{fig:1a}-\ref{fig:1b}.
These images were already shown to have nanostructures with low and high alignment, respectively.
Figure \ref{fig:6} shows length-weighted orientational order parameter plots for each of these images.
Focusing on the single order parameter metric ($S^{w}_{2}$), there is a clear difference in its magnitude for the poorly aligned nanostructure image (Figure \ref{fig:1a}) compared to those for the highly aligned nanostructure images (Figures \ref{fig:1b}-\ref{fig:1c}).
However, for the two highly aligned nanostructure images, the difference in $S^{w}_{2}$ is very small.
This implies that the single order parameter quantification of orientational order is not suitable for distinguishing between different highly aligned samples.
Taking into account multiple orientational order parameters enables this comparison, with the caveat that the orientational order parameters and ODF quantify alignment only.
Other metrics such as nanostructure shape or morphology could be included in the comparison, but would require the introduction of additional metrics.

A simple way to compare the ODFs from each image is through treating the set of orientational order parameters (for each image) as a vector $\bm{S}^{w} = [S^{w}_{2n}]$ and compute the Euclidean distance between vectors from pairs of images.
This value can be interpreted as a similarity metric, the smaller its value the more similar the orientational character of the nanostructures shown in the pair of images is, the larger the more different.
Table \ref{tab:distance} shows the similarity metric values resulting from comparing each of the three SEM images for both the single and multiple order parameter cases.
Both the single and multiple order parameter similarity metrics are found to result in large values comparing the poorly aligned image to both highly aligned images.
While this is correct in both cases, when comparing the highly aligned images to each other the single order parameter similarity metric is very small, which incorrectly implies that these images have very similar alignment characteristics.
The multiple order parameter similarity metric performs well for all cases, indicating that the poorly aligned sample is less similar to the aligned samples and the aligned samples are similar but distinct.

\begin{table}
\caption{Similarity metric values from the comparison of SEM images in Figure \ref{fig:1}.}\label{tab:distance}
\centering
\begin{tabular}{| c | c | c | c|}
\br
 & Figures \ref{fig:1a},\ref{fig:1b} & Figures \ref{fig:1a},\ref{fig:1c} & Figures \ref{fig:1b},\ref{fig:1c} \\
\mr
$|S_{2}^{w} - S^{w'}_{2}|$ & 0.65 & 0.68 & 0.031 \\
$||\bm{S}^{w} - \bm{S}^{w'}||$ & 1.5 & 2.2 & 1.0 \\
\br
\end{tabular}
\end{table}

\section{Conclusions}

Both alignment quantification theory and an image processing method were presented and applied to microscopy images of 1D nanostructures on surfaces.
An appropriate set of two-dimensional orientational order parameters were derived which enable reconstruction of the orientational distribution function (ODF) with which nanostructure alignment is rigorously quantified.
The use of high-order orientional order parameters is shown to be necessary for quantification of highly aligned nanostructures, where past single parameter methods are shown to be insufficient.
Additionally, an image processing method based on mathematical morphology operations is presented which is robust in the presence of measurement uncertainty and nanostructure overlap.

The results of this work provide researchers in nanoscience and nanotechnology with a robust method for the determination of structure/property relationships where alignment of one-dimensional nanostructures is significant.
Subsequently, a fully documented open-source implementation of the method is provided for general use\footnote{The image processing code developed and used in this work is provided in the supplementary information under an open-source license.}.

\section*{Acknowledgements}

This research was supported by the Natural Sciences and Engineering Research Council of Canada (NSERC).

\section*{References}

\bibliographystyle{unsrt}
\bibliography{references,self_assembly,computational} 

\end{document}